\documentclass{article} % For LaTeX2e
\usepackage[T1]{fontenc}
\PassOptionsToPackage{bookmarks=true}{hyperref}
\usepackage{colm2024_conference}

\usepackage{booktabs}
\usepackage{graphicx}
\usepackage{enumitem}
\usepackage{wrapfig}
\usepackage{algorithm}
\usepackage{algpseudocode}
\usepackage{natbib}
\usepackage{makecell}
\usepackage{array}
\usepackage{amsmath}
\usepackage{amssymb}
\usepackage{amsfonts}
\usepackage{multirow}
\usepackage{verbatim}
\usepackage{caption}
\usepackage{longtable}
\usepackage{supertabular}
\usepackage{hyperref}
\hypersetup{bookmarks=true, bookmarksopen=true}

\usepackage{pifont}
\usepackage{afterpage}
\usepackage{tablefootnote}
\usepackage{xspace}
\usepackage{textcomp}
\usepackage{lscape}
\usepackage{siunitx}
\usepackage{listings}
\usepackage{xcolor}
\usepackage{adjustbox}

\definecolor{dt}{gray}{0.7}
\definecolor{leum-blue}{RGB}{30,90,200}
\colorlet{leum-blue-alpha}{leum-blue!38}

\newcommand{\leumvleight}{{Leum-VL-8B}\xspace}

\usepackage{bbding}
\usepackage{fontawesome}
\usepackage{scrextend}
\usepackage{latexsym}
\usepackage{microtype}
\definecolor{mydarkblue}{rgb}{0,0.08,0.45}
\definecolor{citecolor}{HTML}{0071BC}
\usepackage{url}
\usepackage{nicefrac}
\usepackage{changepage}
\usepackage{xargs}
\usepackage{wrapfig,lipsum,booktabs}
\usepackage{subcaption}
\usepackage{endnotes}

\usepackage{pgfplots}
\usetikzlibrary{pgfplots.groupplots}
\pgfplotsset{compat=1.3}
\usepackage{tikz}
\usetikzlibrary{patterns}

\usepackage[most]{tcolorbox}
\usepackage{fvextra}
\usepackage[capitalize,noabbrev]{cleveref}
\crefname{section}{Section}{\S\S}
\Crefname{section}{Section}{\S\S}
\crefname{table}{Table}{Tables}
\crefname{figure}{Figure}{Figures}
\crefname{algorithm}{Algorithm}{}
\crefname{equation}{eq.}{}
\crefname{appendix}{Appendix}{}
\crefformat{section}{Section #2#1#3}
\usepackage{multicol}
\usepackage{fancyvrb}
\usepackage{tcolorbox}
\newsavebox{\myverbcontent}
\usepackage{titlesec}
\titleformat*{\section}{\large\bfseries}

\usepackage{nicematrix}
\usepackage{arydshln}

\makeatletter
\DeclareRobustCommand\onedot{\futurelet\@let@token\@onedot}
\def\@onedot{\ifx\@let@token.\else.\null\fi\xspace}
\def\etc{\emph{etc}\onedot}

\makeatother

\title{Leum-VL Technical Report}

\author{
\bf Yuxuan He$^{*}$,
\bf Chaiming Huang$^{*\dagger}$,
\bf Yifan Wu$^{*}$,
\bf Hongjun Wang$^{*}$, \\
\bf Chenkui Shen,
\bf Jifan Zhang,
\bf Long Li \\[0.3em]
Hainan Sihe Data Technology Co., Ltd. \\[0.3em]
}

\begin{document}

\maketitle
% Author footnote: placed at bottom of first page without auto-generated marker
{\renewcommand{\thefootnote}{}\footnotetext{$^{*}$Core contribution.\quad $^{\dagger}$Corresponding author: \texttt{chaiming@sihe.ai}}}

\begin{abstract}
A short video succeeds not simply because of what it shows, but because of how it schedules attention---yet current multimodal models lack the structural grammar to parse or produce this organization.
Existing models can describe scenes, answer event-centric questions, and read on-screen text, but they are far less reliable at identifying timeline-grounded units such as hooks, cut rationales, shot-induced tension, and platform-facing packaging cues.

We propose \textbf{SV6D} (Structured Video in Six Dimensions), inspired by professional storyboard practice in film and television production, a representation framework that decomposes internet-native video into six complementary structural dimensions---subject, aesthetics, camera language, editing, narrative, and dissemination---with each label tied to physically observable evidence on the timeline.
We formalize a unified optimization objective over SV6D that combines Hungarian-matched temporal alignment, dimension-wise semantic label distance, and quality regularization.
Building on this framework, we present \leumvleight, an 8B video-language model that realizes the SV6D objective through an expert-driven post-training pipeline, further refined through verifiable reinforcement learning on perception-oriented tasks.

\leumvleight achieves 70.8 on VideoMME (w/o subtitles), 70.0 on MVBench, and 61.6 on MotionBench, while remaining competitive on general multimodal evaluations such as MMBench-EN.
We also construct FeedBench, a benchmark for structure-sensitive short-video understanding.
Our results indicate that the missing layer in video AI is not pixel generation but structural representation: grounded on the timeline, linked to visible evidence, and directly consumable by downstream workflows such as editing, retrieval, recommendation, and generation control, including text-heavy internet video formats with overlays and image-text layouts.
\end{abstract}

\begin{figure*}[h]
\centering
\includegraphics[width=\linewidth]{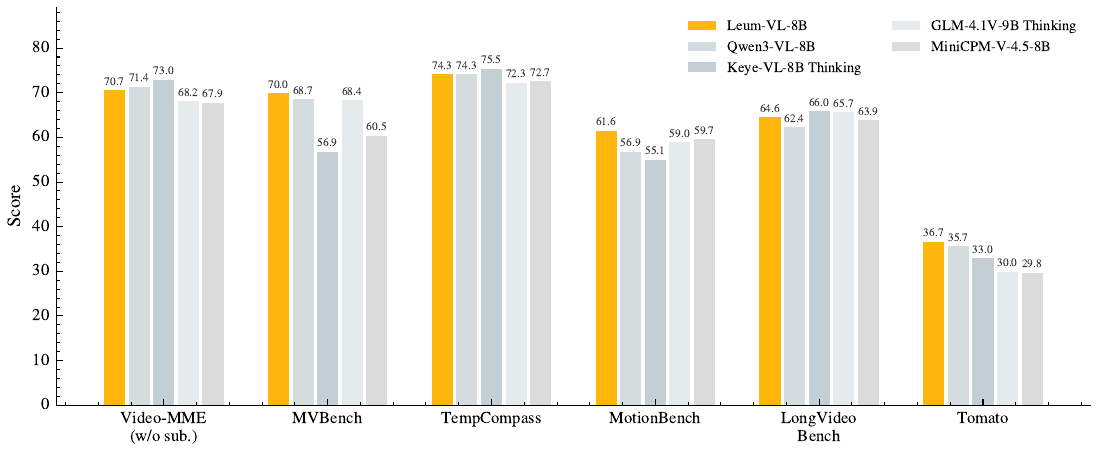}
\caption{Video understanding benchmark comparison of \leumvleight against 8B-scale models.}
\label{fig:benchmark_comparison}
\end{figure*}

\newpage

% ==========================================
\section{Introduction}
\label{sec:intro}
A short video succeeds not simply because of what it shows, but because of how it schedules attention: the opening hook, the rhythm created by shot-size transitions, the release of tension through editing, the guidance provided by subtitles and stickers.
These are not decorative choices---they constitute the \textbf{structural grammar} of video, the layer that professional directors and editors manipulate every day but that current video AI almost entirely ignores.

Existing multimodal models can describe what appears in a frame, answer event-centric questions, and recognize on-screen text, but when faced with structural questions---\textbf{Why does the cut happen here? What narrative function does this segment serve? What dissemination strategy does this shot choice support?}---their answers are often generic platitudes or plausible-sounding ``structural hallucinations'' that cannot be traced back to any concrete evidence on the timeline.
We argue that \textbf{the core missing piece is not larger models or more data, but an explicitly manipulable grammar layer for visual narrative}.

This grammar layer is not a new invention.
In professional film and television production, the \textbf{storyboard} has long played this role: it is the central coordination document for crews of hundreds, where a single storyboard entry simultaneously encodes subject blocking, camera setup, aesthetic intent, editing logic, narrative function, and audience-facing strategy.
These are not six independent annotation tasks but six facets of a single creative specification---together they answer ``why this shot is composed this way.''
Freytag formalized the dramatic tension arc in narrative structure as early as 1863~\cite{freytag1863}; the storyboard extends this principle to the full production stack.
Yet \textbf{no prior work has systematically formalized the storyboard as a computable, trainable, machine-consumable representation}.
This is precisely what we do.

This gap is especially pronounced in \textbf{internet-native short videos}.
Unlike feature films, short videos compress setup, progression, and climax into a few seconds; success depends not only on content but on how information is released and packaged over time---opening tension design, editing rhythm, subtitle overlays, persona markers, interaction prompts.
This is even more evident in text-heavy internet formats, where stickers, embedded UI text, and image-text carousel layouts carry much of the semantic load.

Current training and evaluation paradigms only partially cover these needs.
Dense captioning covers semantics but not shot grammar or narrative role; shot boundary detection captures physical cuts but not their function; aesthetic assessment, narrative labeling, and dissemination analysis are often treated as isolated subtasks rather than coordinated layers on a shared timeline.
The result is that model outputs appear reasonable but are difficult to audit or directly consume downstream.
\textbf{We argue that a model has not truly understood a video unless it can anchor its claims on the timeline and link them to observable evidence.}

We propose \textbf{SV6D} (Structured Video in Six Dimensions), a representation framework that formalizes internet-video understanding as timeline-grounded structural prediction.
SV6D decomposes video into six complementary structural dimensions---\textbf{subject}, \textbf{aesthetics}, \textbf{camera language}, \textbf{editing}, \textbf{narrative}, and \textbf{observable dissemination strategy}---with each label anchored on the timeline and linked to physically observable evidence.
The first five dimensions characterize on-screen content and visual-temporal organization; the sixth focuses on visible, platform-facing packaging signals (e.g., opening tension design, persona markers, interaction prompts) rather than latent dissemination effects.
We define a unified optimization objective over SV6D that combines Hungarian-matched temporal alignment, dimension-wise semantic label distance, and quality regularization.

Building on this framework, we train \textbf{\leumvleight}, an 8B video-language model that realizes the SV6D objective through an expert-driven post-training pipeline combining supervised fine-tuning and verifiable reinforcement learning on perception-oriented tasks.
We also construct \textbf{FeedBench}, a benchmark for structure-sensitive short-video understanding.
Experiments show that \leumvleight achieves strong performance on video understanding evaluations (VideoMME 70.8, MVBench 70.0, MotionBench 61.6) while remaining competitive on general multimodal benchmarks.

Our contributions are fourfold:
\begin{enumerate}
  \item We formulate internet-video understanding as \textbf{timeline-grounded structural parsing}, shifting the objective from free-form commentary to timestamp-aligned, machine-consumable representation.
  \item We propose \textbf{SV6D}, a six-dimensional schema \textbf{inspired by professional storyboard practice} that formalizes the storyboard as a trainable representation, with a unified optimization objective combining temporal alignment, dimension-wise label matching, and quality regularization.
  \item We develop \textbf{\leumvleight} and construct \textbf{FeedBench}, demonstrating strong gains on video-centric and structure-sensitive tasks while maintaining competitive general multimodal performance.
  \item We demonstrate a highly cost-effective post-training paradigm: \leumvleight achieves state-of-the-art structural parsing using only \textbf{4,800 GPU hours}.
\end{enumerate}

% ==========================================
\section{Structured Video in Six Dimensions (SV6D)}
\label{sec:sv6d}
% SV6D: Structured Video in Six Dimensions
% 形式化定义六维视频理解框架、语义镜头、优化目标

This section formalizes \textbf{SV6D}, the representation framework that underpins our approach to cinematic video understanding.
We define \emph{semantic shots} as the minimal composable units of visual storytelling, introduce a \emph{six-dimensional label schema} grounded in film theory and production practice, and derive a unified optimization objective that connects temporal alignment, structural label matching, and output quality regularization.

%----------------------------------------------------------
\subsection{Video as a Multimodal Time Series}
\label{sec:sv6d:video}

Let $V$ denote a video. Its observation comprises three components:

\begin{itemize}[nosep,leftmargin=1.5em]
  \item \textbf{Visual stream}: a frame sequence $X = \{x(t) \mid t \in [0, T]\}$, or in discrete form $\{x_t\}_{t=1}^{T_f}$.
  \item \textbf{Audio stream} (optional): $A = \{a(t)\}$.
  \item \textbf{Metadata}: $M$ (frame rate, resolution, aspect ratio, platform type, whether AIGC-generated, \etc).
\end{itemize}

Our focus is not on per-frame recognition (``what is present''), but rather on \emph{how these observations are organized into composable narrative units, and what intentions underlie those organizational decisions}.

%----------------------------------------------------------
\subsection{Temporal Primitives: Semantic Shots and Skeleton Segments}
\label{sec:sv6d:primitives}

\subsubsection{Semantic Shots (Shot Tokens)}

Traditional shot boundary detection defines a shot as the frame span between two physical cuts.
We adopt a definition closer to creative practice and understanding:

\begin{tcolorbox}[colback=gray!5, colframe=gray!60, title={\textbf{Definition 1} (Semantic Shot)}]
A \emph{semantic shot} $S_i = [t_i^s,\, t_i^e]$ is a maximal contiguous interval on the timeline such that a dominant \emph{creative continuity unit}
\begin{equation}
  u \;\in\; \{\,\text{motion unit},\;\text{emotional unit},\;\text{narrative-action unit}\,\}
\end{equation}
remains unbroken within the interval, and the boundary $t_i^e$ (equivalently $t_{i+1}^s$) corresponds to the \emph{minimal} structural change point that disrupts that dominant unit.
\end{tcolorbox}

The semantic shot sequence $\{S_i\}_{i=1}^{N}$ forms a \emph{partition} of $[0, T]$: non-overlapping and gap-free.
Crucially, shot boundaries mark \emph{creative-structural} transitions, which need not coincide with pixel-level discontinuities.

\subsubsection{Shot-Aligned Skeleton Segments (Discourse Structure)}

Short-form videos and advertisements are typically organized first by \emph{functional segments} (e.g.\ exposition, rising action, climax), with shot-level cinematography serving as the carrier.
We elevate the skeleton to a first-class citizen, but operationalize it as \emph{taxonomy-controlled grouping over the shot token sequence} rather than free-form span annotation on the continuous timeline.

\begin{tcolorbox}[colback=gray!5, colframe=gray!60, title={\textbf{Definition 2} (Shot-Aligned Skeleton)}]
Given a semantic shot sequence $\{S_i\}_{i=1}^{N}$, select a discourse skeleton taxonomy $\tau$ (i.e.\ a \texttt{skeleton\_type}) with segment-function label set $\mathcal{R}_\tau$.
Assign each shot a segment-function label:
\begin{equation}
  r_i \;\in\; \mathcal{R}_\tau, \qquad i = 1, \dots, N.
\end{equation}
\end{tcolorbox}

For the default \texttt{dramatic\_arc} taxonomy, following Freytag's pyramid~\cite{freytag1863}:
\begin{equation}
  \mathcal{R}_{\texttt{dramatic\_arc}} = \{\,\text{exposition},\;\text{rising action},\;\text{climax},\;\text{falling action},\;\text{d\'{e}nouement},\;\text{other}\,\}.
\end{equation}
Other taxonomies (e.g.\ \texttt{three\_act}, \texttt{ki\_sh\={o}\_ten\_ketsu}, \texttt{tutorial}) define their own $\mathcal{R}_\tau$.

The skeleton segment sequence $\{G_j\}_{j=1}^{K}$ is obtained by \emph{maximal contiguous merging} of the shot-level label sequence $\{r_i\}_{i=1}^{N}$.
Each segment $G_j$ corresponds to a contiguous shot range $[i_j^s,\, i_j^e]$ satisfying:

\begin{enumerate}[nosep,leftmargin=1.5em]
  \item \textbf{Coverage \& non-overlap} (shot-index partition): $i_1^s = 1$, $i_K^e = N$, and $i_j^e + 1 = i_{j+1}^s$ for all $j$.
  \item \textbf{Intra-segment homogeneity}: $r_i = \mathrm{seg\_type}(G_j)$ for all $i \in [i_j^s,\, i_j^e]$.
  \item \textbf{Maximality}: adjacent segments have distinct $\mathrm{seg\_type}$ (otherwise they would be merged).
  \item \textbf{Time boundaries as derived quantities}: $\tau_j^s = t_{i_j^s}^s$, \; $\tau_j^e = t_{i_j^e}^e$.
\end{enumerate}

%----------------------------------------------------------
\subsection{Six-Dimensional Schema}
\label{sec:sv6d:schema}

Each semantic shot $S_i$ is annotated with labels from six complementary structural dimensions, denoted $\mathbf{l}_i = (l_{i,1}, \dots, l_{i,6})$ where $l_{i,k} \in \mathcal{L}_k$.
Together, these dimensions capture distinct layers of how a video is organized and communicated over time.

\textbf{Dimension 1: Camera Language} ($\mathcal{L}_1$).
This dimension characterizes how a scene is physically framed and captured, covering shot size, camera position, shooting angle, lens focal length, camera movement, and depth of field.
Each sub-dimension is defined over a closed label space with expert-derived confusion neighborhoods (e.g., \textit{medium close-up} vs.\ \textit{medium shot}, \textit{smooth tracking} vs.\ \textit{pan/tilt}).
Formally, $\mathcal{L}_1 = \mathcal{L}_{\text{shot\_size}} \times \mathcal{L}_{\text{cam\_pos}} \times \mathcal{L}_{\text{angle}} \times \mathcal{L}_{\text{focal}} \times \mathcal{L}_{\text{movement}} \times \mathcal{L}_{\text{dof}}$.

\textbf{Dimension 2: Aesthetics} ($\mathcal{L}_2$).
This dimension captures the visually expressive properties that shape attention and interpretation: light source nature, light direction, light hardness, contrast, saturation, color temperature, key/tone, composition, and visual rhythm.
Labels are grounded in physically observable evidence and span nine sub-dimensions with 41 canonical tags.
Formally, $\mathcal{L}_2 = \mathcal{L}_{\text{light\_src}} \times \mathcal{L}_{\text{light\_dir}} \times \mathcal{L}_{\text{hardness}} \times \mathcal{L}_{\text{contrast}} \times \mathcal{L}_{\text{saturation}} \times \mathcal{L}_{\text{color\_temp}} \times \mathcal{L}_{\text{key}} \times \mathcal{L}_{\text{comp}} \times \mathcal{L}_{\text{rhythm}}$.

\textbf{Dimension 3: Editing} ($\mathcal{L}_3$).
This dimension covers the structural logic of how shots are assembled: editing logic (16 labels including continuity cut, montage, jump cut, match cut, \etc), editing effects (19 labels), and transition type (organized into four groups: base, opacity \& light, motion \& displacement, and distortion \& glitch).
Timestamp-anchored items are used for boundary-sensitive judgments such as cut rationale.
Formally, $\mathcal{L}_3 = \mathcal{L}_{\text{edit\_logic}} \times \mathcal{L}_{\text{edit\_effect}} \times \mathcal{L}_{\text{transition}}$.

\textbf{Dimension 4: Subject Analysis} ($\mathcal{L}_4$).
This dimension characterizes who or what is foregrounded, combining framing type (over-the-shoulder, insert, subjective/POV, \etc) with subject configuration (single-person, two-person, group, none).
Invalid combinations serve as quality-control signals during annotation.
Formally, $\mathcal{L}_4 = \mathcal{L}_{\text{framing}} \times \mathcal{L}_{\text{config}}$.

\textbf{Dimension 5: Narrative} ($\mathcal{L}_5$).
This dimension labels the functional role of each temporal segment in the information-release arc: exposition, rising action, escalation, peripeteia, climax, falling action, and d\'{e}nouement.
It relies heavily on ordered-transition items, as many meaningful narrative units are transitional rather than static.
Formally, $\mathcal{L}_5 = \mathcal{L}_{\text{seg\_func}}$, where $\mathcal{L}_{\text{seg\_func}}$ is coupled with the skeleton taxonomy $\mathcal{R}_\tau$ (Definition~2).

\textbf{Dimension 6: Dissemination} ($\mathcal{L}_6$).
This dimension is restricted to visible, platform-facing packaging cues: retention engine labels (observable tension devices, interaction prompts, engagement cues) and comment alignment tasks.
Crucially, it targets observable strategy rather than latent virality or platform outcome.
Formally, $\mathcal{L}_6 = \mathcal{L}_{\text{retention}} \times \mathcal{L}_{\text{comment\_align}}$.

Together, these six dimensions form a shared timeline-grounded schema.
Subject, aesthetics, and camera language characterize what appears on screen and how it is visually constructed.
Editing and narrative capture how information is segmented, sequenced, and made consequential over time.
Dissemination captures how content is packaged for platform-native circulation.
All six layers are defined on a common timeline, enabling a unified structural parse rather than disconnected subtask outputs.

These six dimensions form a \textbf{joint structured annotation} over each semantic shot, analogous to a professional storyboard entry that simultaneously specifies subject blocking, camera setup, aesthetic intent, editing logic, narrative function, and dissemination strategy.
The SV6D parse of a video is therefore a \textbf{single structured prediction task} producing a complete shot-aligned document:
\begin{equation}
\label{eq:dobs}
  D^{\mathrm{obs}} \;=\; \bigl\{\, \bigl(S_i,\; \mathbf{l}_i,\; r_i \bigr) \,\bigr\}_{i=1}^{N}
\end{equation}
rather than six independent classification problems.
All six layers share a common timeline, enabling cross-dimensional causal reasoning---for example, why a particular shot size is chosen given the narrative function and dissemination intent of the enclosing segment.
We refer to $D^{\mathrm{obs}}$ as the \emph{observed structural document}; this symbol is reused in the data pipeline (\cref{sec:data_6d}) and training objective (\cref{sec:rlvr}).

%----------------------------------------------------------
\subsection{SV6D Optimization Objective}
\label{sec:sv6d:objective}

We now formalize the optimization objective for training a model to produce SV6D-compliant structural parses.
Given a video $V$, the model predicts a shot sequence $\hat{\mathbf{S}} = \{\hat{S}_i\}_{i=1}^{\hat{N}}$ with per-shot label vectors $\hat{\mathbf{l}}_i = (\hat{l}_{i,1}, \dots, \hat{l}_{i,6})$.
The ground-truth annotation is $\mathbf{S}^{*} = \{S_j^{*}\}_{j=1}^{N^{*}}$ with labels $\mathbf{l}_j^{*}$.
The SV6D loss decomposes into three terms:

\begin{equation}
\label{eq:sv6d_loss}
  \mathcal{L}_{\mathrm{SV6D}} \;=\; \underbrace{\mathcal{L}_{\mathrm{align}}}_{\text{temporal alignment}} \;+\; \underbrace{\mathcal{L}_{\mathrm{struct}}}_{\text{structural matching}} \;+\; \underbrace{\mathcal{L}_{\mathrm{reg}}}_{\text{quality regularization}}
\end{equation}

\subsubsection{Temporal Alignment via Hungarian Matching}

Since the predicted and ground-truth shot sequences may differ in cardinality ($\hat{N} \neq N^{*}$), we first establish a correspondence via optimal bipartite matching~\cite{kuhn1955hungarian}.
Define the pairwise cost matrix $C \in \mathbb{R}^{\hat{N} \times N^{*}}$:

\begin{equation}
\label{eq:cost_matrix}
  C_{ij} \;=\; \alpha\,\bigl(1 - \mathrm{IoU}(\hat{S}_i,\, S_j^{*})\bigr) \;+\; (1-\alpha)\,\Delta_{\mathrm{label}}(\hat{\mathbf{l}}_i,\, \mathbf{l}_j^{*})
\end{equation}

where $\mathrm{IoU}(\hat{S}_i, S_j^{*}) = \frac{|\hat{S}_i \cap S_j^{*}|}{|\hat{S}_i \cup S_j^{*}|}$ is the temporal intersection-over-union, $\Delta_{\mathrm{label}}$ is the aggregate label distance defined below, and $\alpha \in (0,1)$ balances temporal and structural costs.

The Hungarian algorithm~\cite{kuhn1955hungarian} yields the optimal assignment $\sigma^{*}$:

\begin{equation}
\label{eq:hungarian}
  \sigma^{*} \;=\; \underset{\sigma \in \Pi(\hat{N},\, N^{*})}{\arg\min} \sum_{(i,j) \in \sigma} C_{ij}
\end{equation}

where $\Pi(\hat{N}, N^{*})$ denotes the set of feasible partial bijections (allowing unmatched shots when $\hat{N} \neq N^{*}$).
Let $\mathcal{M} = \{(i, \sigma^{*}(i))\}$ denote the matched pairs.
The temporal alignment loss is:

\begin{equation}
\label{eq:align_loss}
  \mathcal{L}_{\mathrm{align}} \;=\; \frac{1}{|\mathcal{M}|}\sum_{(i,j)\in\mathcal{M}} \bigl(1 - \mathrm{IoU}(\hat{S}_i,\, S_j^{*})\bigr) \;+\; \beta\,\frac{|\hat{N} - N^{*}|}{\max(\hat{N},\, N^{*})}
\end{equation}

where the second term penalizes cardinality mismatch with coefficient $\beta$.

\subsubsection{Dimension-Wise Structural Matching}

For each matched pair $(i, j) \in \mathcal{M}$ and each dimension $k \in \{1, \dots, 6\}$, we define a label distance function $d_k: \mathcal{L}_k \times \mathcal{L}_k \to [0, 1]$ that reflects the semantic proximity between labels.
Critically, $d_k$ is \emph{not} a binary indicator: labels that are semantically close (e.g., \textit{medium shot} vs.\ \textit{medium close-up} in camera language) incur a smaller penalty than labels that are far apart (e.g., \textit{extreme long shot} vs.\ \textit{extreme close-up}).

Each dimension $k$ is equipped with a domain-specific metric derived from the label taxonomy.
For dimensions with ordinal structure (e.g., shot size), $d_k$ is defined by the normalized rank distance on the label hierarchy.
For dimensions with categorical structure (e.g., transition type), $d_k$ is derived from expert-defined confusion neighborhoods, where the distance between two labels corresponds to the shortest path on the confusion graph.

The structural matching loss aggregates over all dimensions with learned weights $w_k > 0$:

\begin{equation}
\label{eq:struct_loss}
  \mathcal{L}_{\mathrm{struct}} \;=\; \sum_{k=1}^{6} w_k \cdot \frac{1}{|\mathcal{M}|} \sum_{(i,j)\in\mathcal{M}} d_k\bigl(\hat{l}_{i,k},\; l_{j,k}^{*}\bigr)
\end{equation}

subject to $\sum_{k=1}^{6} w_k = 1$.
The weights encode the relative importance of each dimension and are tuned on a held-out validation set.

\subsubsection{Quality Regularization}

Beyond structural correctness, the model output must satisfy quality constraints on terminology, completeness, and format.
We introduce a regularization term analogous to the KL penalty in RLHF:

\begin{equation}
\label{eq:reg_loss}
  \mathcal{L}_{\mathrm{reg}} \;=\; \lambda_p \cdot \mathcal{R}_{\mathrm{prof}}(\hat{y}) \;+\; \lambda_c \cdot \mathcal{R}_{\mathrm{comp}}(\hat{y}) \;+\; \lambda_f \cdot \mathcal{R}_{\mathrm{form}}(\hat{y})
\end{equation}

where $\hat{y}$ denotes the full model output and:

\begin{itemize}[nosep,leftmargin=1.5em]
  \item $\mathcal{R}_{\mathrm{prof}}$ penalizes out-of-vocabulary label predictions.
    Let $\mathcal{V}_k$ be the closed canonical label set for dimension $k$ (e.g., the 38 camera-language tags or 41 aesthetics tags defined in \cref{sec:sv6d:schema}).
    $\mathcal{R}_{\mathrm{prof}}$ counts the fraction of predicted labels that do not string-match any element in $\bigcup_k \mathcal{V}_k$, so that a model producing ``medium close'' instead of the canonical ``medium close-up'' is penalized.
  \item $\mathcal{R}_{\mathrm{comp}}$ penalizes missing dimensions.
    For each of the six dimensions, a binary indicator checks whether the output contains at least one label for that dimension; $\mathcal{R}_{\mathrm{comp}}$ is the number of absent dimensions divided by 6.
    A prediction that omits editing labels entirely receives $\mathcal{R}_{\mathrm{comp}} \geq 1/6$ regardless of how accurate the remaining dimensions are.
  \item $\mathcal{R}_{\mathrm{form}}$ penalizes structural format violations detectable by deterministic parsing: invalid JSON syntax, missing required fields (shot boundaries, dimension keys), malformed timestamps (e.g., negative values or $t^s > t^e$), and type errors (string where a float is expected).
    Each violation type contributes a binary penalty, and $\mathcal{R}_{\mathrm{form}}$ is the fraction of violated checks.
\end{itemize}

The coefficients $\lambda_p, \lambda_c, \lambda_f > 0$ control the strength of each regularizer.

\subsubsection{Summary}

The complete SV6D objective (\cref{eq:sv6d_loss}) provides a principled, differentiable measure of structural video understanding quality.
In practice, this objective is realized through two complementary training stages:
Supervised fine-tuning (SFT) minimizes a proxy of $\mathcal{L}_{\mathrm{SV6D}}$ via structured supervision (\cref{sec:sft}), while GRPO directly optimizes a reward function derived from $\mathcal{L}_{\mathrm{SV6D}}$, where the reward for each rollout is $r = 1 - \mathcal{L}_{\mathrm{SV6D}}$ (\cref{sec:rlvr}).
The Hungarian matching and dimension-wise distance functions described above correspond directly to the IoU and label-similarity components of the GRPO reward (Table~\ref{tab:reward_functions}).

% ==========================================
\section{Training Data and Synthesis}
\label{sec:data}

\begin{figure*}[t]
\centering
\includegraphics[width=\linewidth]{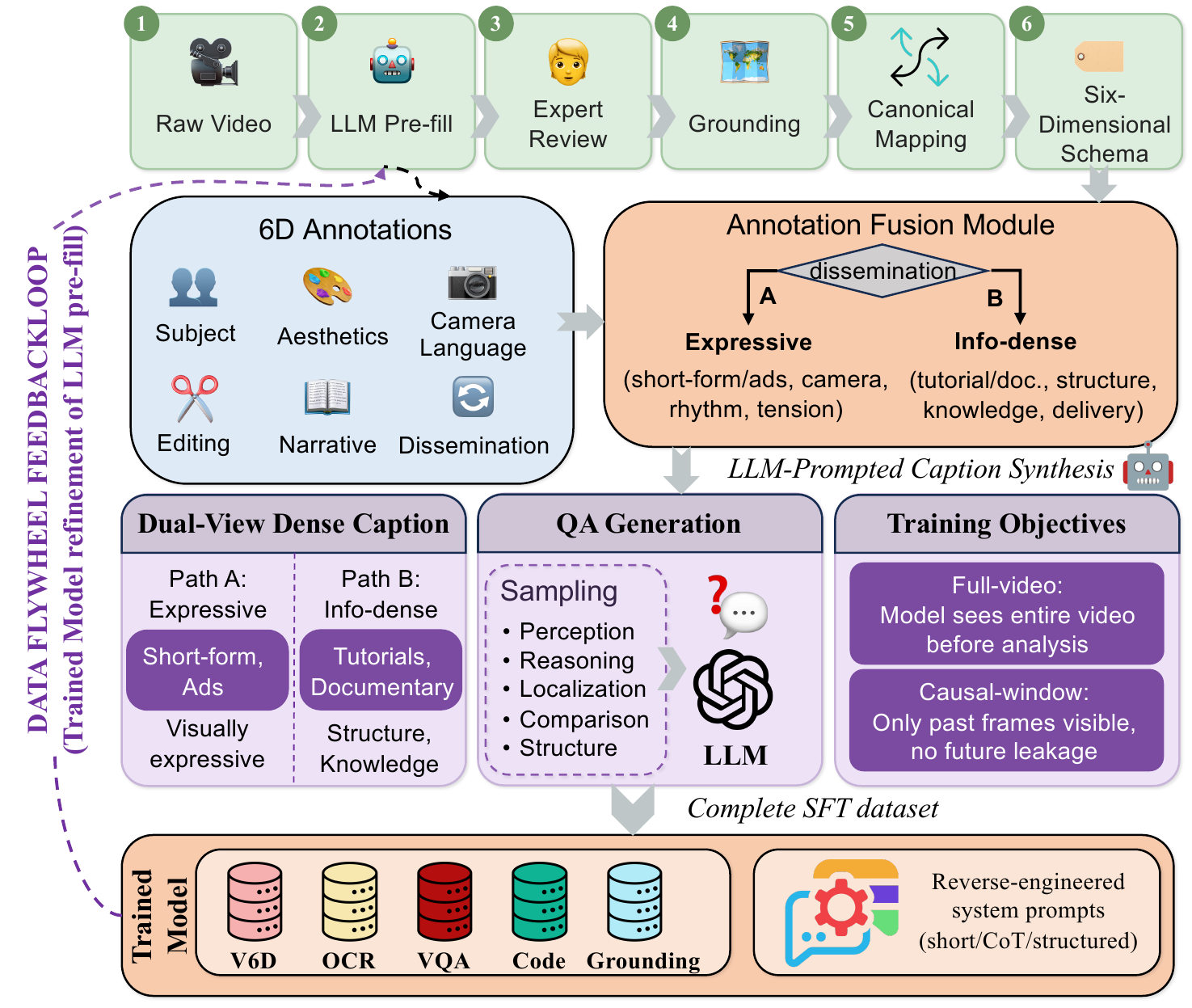}
\caption{Overview of the data synthesis pipeline. Starting from raw internet video, the pipeline proceeds through shot segmentation, LoRA-based structural annotation, scorer-based quality filtering, LLM-assisted structural extension, and dense report generation, producing the SV6D SFT training corpus.}
\label{fig:pipeline}
\end{figure*}

\subsection{SV6D Annotation}
\label{sec:data_6d}

The SV6D schema defined in \cref{sec:sv6d:schema} is instantiated through a combination of expert annotation and automated synthesis.

\subsubsection{Expert Annotation}

The SV6D schema is instantiated through expert annotation by specialists spanning directing, cinematography, screenwriting, planning, sound, and internet operations.
Each structural label is tied to physically observable evidence on the timeline before being mapped to a canonical tag.
Evidence may come from the visual stream, time-aligned speech transcription, or on-screen text, but in all cases the target remains auditable against the source video.
This prevents labels from degenerating into unconstrained interpretation or post-hoc rationalization.

The annotation process produces structured video decomposition reports covering all six dimensions.
These reports serve as the ground-truth backbone for both the FeedBench evaluation benchmark (Section~\ref{sec:benchmark}) and the SFT data synthesis pipeline described below.

\subsubsection{Automated Annotation and Quality Filtering}

To scale beyond fully manual annotation, we use a semi-automatic structural labeling pipeline aligned with the expert-defined schema.
The pipeline first proposes candidate timeline-grounded annotations for unannotated short videos and then applies multi-stage quality control to remove low-confidence or schema-inconsistent cases.
Rather than relying on a single-pass automatic labeler, we combine confidence-based filtering, cross-field consistency checks, and report-level validation so that retained annotations remain compatible with the same six-dimensional representation used by expert annotators.

For videos that pass filtering, the retained annotations are consolidated into complete structural decomposition reports and converted into complementary supervision signals, including structured outputs, dense descriptions, and automatically checkable QA instances.
This design allows expert-defined structure to be expanded at scale while preserving auditability, temporal grounding, and label consistency across heterogeneous short-video formats.
The pipeline also supports incremental large-scale processing and selective reprocessing as annotation rules and models improve.

\subsubsection{QA Data Synthesis}

Building on the structured decomposition reports, we synthesize large-scale SFT question--answer pairs using a programmatic QA generation framework grounded in the SV6D schema.
For each video, a capable LLM is prompted with the six-dimensional decomposition report and a task-specific instruction template to generate diverse QA pairs.
The synthesized data spans four complementary task families:

\begin{itemize}[nosep]
  \item \textbf{Attribute-specific QA.} Given a temporal segment, the model is asked to identify one or more structural labels (e.g., shot size, lighting direction, editing logic). Templates range from single-dimension queries to multi-dimension ``snapshot'' requests covering all six dimensions simultaneously.
  \item \textbf{Temporal grounding and retrieval.} The model must locate all temporal segments matching a given structural label, or identify which label applies at a queried timestamp. This enforces strong timestamp--attribute binding.
  \item \textbf{Abductive reasoning and evidence explanation.} Given an observed structural decision (e.g., a specific cut or shot-size transition), the model must infer the underlying directorial intent and cite observable evidence from the timeline.
  \item \textbf{Error correction.} The model is presented with a deliberately perturbed decomposition report containing incorrect labels, and must identify and correct the errors with justification.
\end{itemize}

Each task family is instantiated across all applicable SV6D dimensions, producing approximately \textbf{2.5M QA pairs} in total.
Multiple-choice variants are also generated for discriminative evaluation.

\subsection{SFT Dataset Composition}
\label{sec:data_sft_composition}

To equip our model with expert-level cinematic understanding while preserving broad multimodal competence, we construct a large-scale SFT dataset comprising approximately \textbf{1M samples}, organized into two complementary pillars: domain-specific data targeting cinematic and video production expertise, and general-purpose data ensuring robust foundational capabilities.

\textbf{Domain-Specific Video Understanding.}
The core of our training data consists of proprietary video annotations produced by professional directors and screenwriters, covering dense captions, structured QA synthesized under the SV6D framework (including multiple-choice, natural-language dialogue, and video caption summarization), ASR-derived tasks, and in-video OCR samples.
We apply balanced sampling across video categories and SV6D label dimensions to prevent category imbalance.

\textbf{OCR.}
We curate OCR data spanning five complementary tasks from established public benchmarks:
text recognition~\citep{mishra2012iiit5k,karatzas2013icdar,karatzas2015icdar,wiedmann2025finevision},
document VQA~\citep{mathew2021docvqa},
scene text VQA~\citep{singh2019textvqa,biten2019stvqa},
key information extraction~\citep{park2019cord,huang2019sroie,jaume2019funsd},
and handwritten math recognition~\citep{yuan2022hme100k,mouchere2013crohme,mathwriting2024,onethinker2024}.

\textbf{Visual Question Answering \& Reasoning.}
This subset incorporates general visual instruction following~\citep{liu2023llava}, chain-of-thought visual reasoning~\citep{xu2024llavacot}, chart understanding~\citep{masry-etal-2022-chartqa}, abstract diagram reasoning~\citep{lu2021iconqa}, multi-step reasoning~\citep{chia-etal-2024-puzzlevqa,lu2022learn}, diverse multimodal tasks~\citep{laurencon2024matters}, and UI screen understanding~\citep{wiedmann2025finevision}.

\textbf{Video Understanding.}
LLaVA-Video-178K~\citep{zhang2024videoinstructiontuningsynthetic} provides both multiple-choice and open-ended video QA covering academic-style visual reasoning.
FineVideo~\citep{Farre2024FineVideo} contributes fine-grained video QA pairs with detailed temporal annotations.

\textbf{Grounding \& Counting.}
We incorporate data from PixMo~\citep{deitke2024molmo} spanning object counting, point-based localization, and spatial grounding tasks.

\textbf{Code Generation.}
We combine Design2Code~\citep{si2024design2code}, WebSight~\citep{laurencon2024websight}, and VisCode-200K~\citep{ni2025viscoder} for UI-to-code and visualization code generation.

\textbf{Text-Only Reasoning \& Instruction Following.}
To maintain strong linguistic and reasoning capabilities, we incorporate text-only samples spanning mathematical reasoning~\citep{zhou2025megamath,du2025nemotronmath}, general instruction following~\citep{teknium2023openhermes,bai2024coig,kopf2023openassistant}, test-time scaling~\citep{muennighoff2025s1simpletesttimescaling}, complex synthetic reasoning~\citep{zhu2026chimera}, safety alignment~\citep{wildguard2024}, and function calling~\citep{liu2024apigen}.

% ==========================================
\section{FeedBench}
\label{sec:benchmark}

\subsection{Limitations of Existing Benchmarks}

Recent benchmarks such as Video-MME, MVBench, TempCompass, MotionBench, and LongVideoBench have advanced multimodal video evaluation, but they remain fundamentally \textbf{event-centric}: they test what happens in a clip, not how the video is structurally organized over time.
Tasks such as dense captioning, shot boundary detection, aesthetics scoring, and social-media analysis are treated as isolated subtasks rather than coordinated layers on a shared timeline, making it difficult to assess whether a model has formed a unified structural parse.
Furthermore, benchmark outputs are rarely timestamp-aligned or schema-consistent, leaving a gap between benchmark performance and the structured, machine-consumable representations required by downstream applications such as editing assistance, retrieval, and generation control.
These shortcomings are especially pronounced for internet-native short-form video, where camera language, editing rhythm, narrative release, and visible packaging cues are not peripheral style choices but core carriers of meaning.

\subsection{FeedBench Design}

FeedBench evaluates all SV6D structural abilities (Section~\ref{sec:sv6d}) on a shared timeline, asking whether a model can recover a timestamp-aligned, machine-consumable structural parse of how a video is organized over time.

\textbf{Evidence grounding.} Each structural label is tied to physically observable evidence on the timeline before being mapped to a canonical tag. Evidence may come from the visual stream, time-aligned speech transcription, or on-screen text, but in all cases the target remains auditable against the source video.

\textbf{Internet-native scope.} The benchmark is drawn from platform-native internet video complemented by TVC-style commercial short videos, spanning 27 vertical categories. This reflects the diversity of real short-video ecosystems, including information-centric, performance-driven, product-led, and aesthetic-edit formats.

\textbf{Expert-driven annotation.} Labels are constructed by specialists spanning directing, cinematography, screenwriting, planning, and internet operations. Generic crowdsourced labeling is insufficient for the expert-sensitive boundaries that define real annotation difficulty in structure analysis.

\textbf{Text-heavy internet-native formats.} Many real short videos distribute their semantics across captions, stickers, embedded UI text, and image-text layouts. FeedBench explicitly includes these overlay-dominant formats rather than treating them as edge cases.

\textbf{Machine-verifiable evaluation.} Benchmark instances are derived from canonical labels and temporal anchors, enabling reproducible and judge-free assessment. Structure-conditioned external validations (e.g., comment alignment) serve as downstream transfer tests rather than part of the core metric.

\subsection{Task Instantiation from the Structural Schema}

Formally, each FeedBench item is a tuple:
\begin{equation}
\label{eq:feedbench_item}
  q \;=\; \bigl(\tau,\; d,\; \mathcal{L}_d,\; \mathbf{t},\; \mathbf{y}^{*},\; \mathbf{e},\; \phi\bigr)
\end{equation}
where $\tau$ is the temporal anchor (interval $[t_s, t_e]$ for persistent properties, or timestamp $t$ for boundary-sensitive judgments), $d \in \{1, \ldots, 6\}$ indexes the SV6D dimension, $\mathcal{L}_d$ is its canonical label space (\cref{sec:sv6d:schema}), $\mathbf{t}$ is a question template, $\mathbf{y}^{*}$ is the ground-truth label(s), $\mathbf{e}$ is the evidence description, and $\phi \in \{\texttt{single},\, \texttt{multi},\, \texttt{ordered}\}$ is the answer type.

The matching function is:
\begin{equation}
\label{eq:match}
  \mathrm{match}(o,\, \mathbf{y}^{*},\, \phi) \;=\;
  \begin{cases}
    o = \mathbf{y}^{*} & \phi = \texttt{single} \\
    o = \mathbf{y}^{*} \;\text{(set equality)} & \phi = \texttt{multi} \\
    o = \mathbf{y}^{*} \;\text{(sequence equality)} & \phi = \texttt{ordered}
  \end{cases}
\end{equation}

FeedBench instantiates the SV6D schema as closed-vocabulary, judge-free tasks over a shared timeline.
Each item is anchored to either a temporal interval (for persistent properties such as shot framing, lighting, or narrative role) or a timestamp (for boundary-sensitive judgments such as cut rationale).

We use three answer types:
\textbf{Single-label classification} for a single canonical tag;
\textbf{multi-label classification} when multiple tags legitimately co-occur within the same anchor;
\textbf{ordered-transition recognition} for progressions rather than states (e.g., a shot-size transition or staged narrative release).
All candidate answers are drawn exclusively from the canonical label space of the queried dimension---no paraphrased options or label variants are permitted.

Distractor construction is \textbf{confusion-aware}: hard negatives are drawn from expert-defined confusion neighborhoods (adjacent shot scales, visually similar lighting, nearby camera-motion categories, etc.).
For ordered-transition items, distractors additionally include temporal traps such as reversal, local substitution, false stasis, or skipped progression.
The surface form of each item is template-controlled and answer-agnostic: the question stem does not reveal the underlying evidence description or paraphrase the ground-truth label.

\subsection{Benchmark Details}
\label{sec:label_taxonomy}

\textbf{Label taxonomy.} Each of the six dimensions uses a closed vocabulary of canonical tags designed by specialists spanning directing, cinematography, screenwriting, and internet operations; sub-dimensions are listed in Table~\ref{tab:label_taxonomy}.

\textbf{Item generation.} Each item is generated programmatically from a structured configuration; \cref{alg:item_gen} formalizes the procedure.
The seven invariants enforced during generation are stated below in terms of the item tuple (\cref{eq:feedbench_item}):

\textbf{I1 (Label atomicity).}
$\forall\; o \in \mathrm{Options}(q):\; o \in \mathcal{L}_d$.

\textbf{I2 (Format homogeneity).}
\begin{align}
  \phi = \texttt{single} &\implies \mathrm{Options} = \{o_A, o_B, o_C, o_D\} \subset \mathcal{L}_d \notag \\
  \phi = \texttt{multi}  &\implies \mathrm{Options} = \{O_A, O_B, O_C, O_D\},\; O_i \subseteq \mathcal{L}_d \notag \\
  \phi = \texttt{ordered} &\implies \mathrm{Options} = \bigl\{(l_1^A \!\to\! \cdots \!\to\! l_m^A),\, \ldots,\, (l_1^D \!\to\! \cdots \!\to\! l_m^D)\bigr\} \notag
\end{align}
All items include option $E$ = ``none of the above / cannot determine.''

\textbf{I3 (Unique correct answer).}
$\bigl|\{o \in \{A,B,C,D\} : \mathrm{match}(o,\, \mathbf{y}^{*},\, \phi)\}\bigr| = 1$.

\textbf{I4 (Confusion-aware distractors).}
Let $\mathcal{N}(\mathbf{y}^{*},\, d)$ be the confusion neighborhood of $\mathbf{y}^{*}$ in dimension $d$:
$\bigl|\mathrm{Distractors}(q) \cap \mathcal{N}(\mathbf{y}^{*},\, d)\bigr| \geq 1$;
target: $\geq 2$ of 3 distractors from $\mathcal{N}$.

\textbf{I5 (Temporal traps).}
When $\phi = \texttt{ordered}$, distractors include $\geq 2$ trap types from $\{\text{reversal},\, \text{local substitution},\, \text{false stasis}\; (X \!\to\! X),\, \text{skip}\}$.

\textbf{I6 (Answer-agnostic surface).}
Neither $\mathbf{y}^{*}$ nor $\mathbf{e}$ appears in the question stem.

\textbf{I7 (Legality fallback).}
$\mathbf{y}^{*} \notin \mathcal{L}_d \implies$ correct answer $= E$.

\textbf{Evaluation protocol.} Scoring is deterministic and rule-based (LLM judge as fallback for non-conforming output only). All three item types require exact match---no partial credit. Results are reported per-dimension, per-answer-type, and as macro-averaged summary scores; a confusion-aware subset score is additionally reported over hard-confusion items.

Per-dimension accuracy:
\begin{equation}
\label{eq:acc_d}
  \mathrm{Acc}_d \;=\; \frac{1}{|Q_d|} \sum_{q \in Q_d} \mathbb{1}\bigl[\mathrm{match}(\hat{y}_q,\, \mathbf{y}^{*}_q,\, \phi_q)\bigr]
\end{equation}

Macro-averaged score:
\begin{equation}
\label{eq:feedbench_macro}
  \mathrm{FeedBench}_{\mathrm{macro}} \;=\; \frac{1}{|\mathcal{D}|} \sum_{d \in \mathcal{D}} \mathrm{Acc}_d
\end{equation}

Confusion-aware subset score:
\begin{equation}
\label{eq:feedbench_hard}
  \mathrm{FeedBench}_{\mathrm{hard}} \;=\; \frac{1}{|\mathcal{D}|} \sum_{d \in \mathcal{D}} \mathrm{Acc}_d^{\,\mathcal{N}}
\end{equation}
where $\mathrm{Acc}_d^{\,\mathcal{N}}$ is computed only on items whose distractors are drawn from $\mathcal{N}$.

By-answer-type breakdown:
\begin{equation}
\label{eq:acc_phi}
  \mathrm{Acc}_\phi \;=\; \frac{1}{|Q_\phi|} \sum_{q \in Q_\phi} \mathbb{1}\bigl[\mathrm{match}(\hat{y}_q,\, \mathbf{y}^{*}_q,\, \phi)\bigr], \quad \phi \in \{\texttt{single},\, \texttt{multi},\, \texttt{ordered}\}
\end{equation}

\textbf{Composition.} Table~\ref{tab:feedbench_overview} summarizes the full benchmark.

\begin{table*}[t]
\centering
\small
\setlength{\tabcolsep}{5pt}
\caption{Per-dimension label taxonomy of FeedBench. Sub-dimensions are listed for each of the six structural dimensions.}
\label{tab:label_taxonomy}
\begin{tabular}{ll}
\toprule
\textbf{Dimension} & \textbf{Sub-dimensions} \\
\midrule
\multirow{6}{*}{Camera language}
  & Shot size         \\
  & Camera position   \\
  & Shooting angle    \\
  & Lens focal length \\
  & Camera movement   \\
  & Depth of field    \\
\midrule
\multirow{9}{*}{Aesthetics}
  & Light source nature \\
  & Light direction     \\
  & Light hardness      \\
  & Contrast            \\
  & Saturation          \\
  & Color temperature   \\
  & Key/tone            \\
  & Composition         \\
  & Visual rhythm       \\
\midrule
\multirow{2}{*}{Editing}
  & Editing logic    \\
  & Editing effects  \\
\midrule
Subject
  & Framing $\times$ configuration \\
\midrule
\multirow{3}{*}{Narrative}
  & Content structure \\
  & Narrative structure \\
  & Narrative techniques \\
\midrule
\multirow{2}{*}{Dissemination}
  & Retention engine \\
  & Comment alignment \\
\bottomrule
\end{tabular}
\end{table*}

\begin{algorithm}[t]
\caption{FeedBench Item Generation}
\label{alg:item_gen}
\begin{algorithmic}[1]
\Require Configuration $\mathcal{C} = (\tau,\; k,\; \mathcal{L}_k,\; \mathcal{T},\; [t_s, t_e],\; y^{*},\; e,\; q_{\mathrm{type}})$
\Statex \hspace{2.4em} where $\tau$: task ID, $k$: dimension, $\mathcal{L}_k$: label space, $\mathcal{T}$: question template,
\Statex \hspace{2.4em} $[t_s, t_e]$: temporal anchor, $y^{*}$: ground-truth labels, $e$: evidence, $q_{\mathrm{type}} \in \{\textsc{single}, \textsc{multi}, \textsc{ordered}\}$
\Ensure MCQ item $\mathcal{I} = (\mathrm{stem},\; \{o_A, o_B, o_C, o_D, o_E\},\; a^{*})$
\Statex
\State $\mathrm{stem} \gets \mathrm{Render}(\mathcal{T},\; [t_s, t_e])$
    \Comment{Inv.~6: answer-agnostic; $y^{*} \notin \mathrm{stem}$, $e \notin \mathrm{stem}$}
\Statex
\If{$y^{*} \notin \mathcal{L}_k$}
    \Comment{Inv.~7: legality fallback}
    \State $a^{*} \gets o_E$; \Return $\mathcal{I}$
\EndIf
\Statex
\State \textbf{Format options by} $q_{\mathrm{type}}$:
    \Comment{Inv.~2: format homogeneity}
\Statex \hspace{2em} \textsc{single}: each $o_i \in \mathcal{L}_k$ \quad (atomic label)
\Statex \hspace{2em} \textsc{multi}: each $o_i \subseteq \mathcal{L}_k$ \quad (label set)
\Statex \hspace{2em} \textsc{ordered}: each $o_i \in \mathcal{L}_k^{+}$ \quad (label sequence, joined by $\rightarrow$)
\Statex \hspace{2em} $o_E \gets$ ``none of the above / cannot determine''
\Statex
\State \textbf{Assert} $\forall\, o_i,\; \mathrm{atoms}(o_i) \subseteq \mathcal{L}_k$
    \Comment{Inv.~1: label atomicity}
\Statex
\State Place $y^{*}$ at a uniformly random position among $\{A, B, C, D\}$
    \Comment{Inv.~3: unique correct answer}
\Statex
\State \textbf{Select distractors} for the remaining three positions:
    \Comment{Inv.~4: confusion-aware}
\Statex \hspace{2em} $d_{\mathrm{hard}} \gets \mathrm{ConfusionNeighborhood}(y^{*},\; \mathcal{L}_k)$ \quad ($\geq 1$ hard negative)
\Statex \hspace{2em} $d_{\mathrm{rest}} \gets \mathrm{Fallback}(e,\; \mathrm{adjacency},\; \mathrm{editorial})$ \quad (target: 2/3 hard)
\Statex
\If{$q_{\mathrm{type}} = \textsc{ordered}$}
    \Comment{Inv.~5: temporal traps}
    \State \textbf{Assert} $|\mathrm{Traps}(\{o_i\}) \cap \{\text{reversal},\, \text{substitution},\, \text{stasis},\, \text{skip}\}| \geq 2$
\EndIf
\Statex
\State \Return $\mathcal{I} = (\mathrm{stem},\; \{o_A, o_B, o_C, o_D, o_E\},\; a^{*})$
\end{algorithmic}
\end{algorithm}

\begin{table}[t]
\centering
\small
\setlength{\tabcolsep}{5pt}
\begin{tabular}{lcccc}
\toprule
Dimension & Temporal anchor & Task Types & \# Labels & \# Items \\
\midrule
Subject & Interval & Single, Ordered & 8 & 88$^\dagger$ \\
Aesthetics & Interval & Single, Multi, Ordered & 41 & 6{,}645 \\
Camera language & Interval & Single, Multi, Ordered & 38 & 5{,}927 \\
Editing & Interval / Timestamp & Single, Multi, Ordered & 65 & 949 \\
Narrative & Interval & Single, Multi, Ordered & [WIP] & [WIP] \\
Dissemination & Interval & Single, Multi & 40 & 194 \\
\midrule
\multicolumn{4}{l}{\textit{Auxiliary: Comment alignment}} & 912 \\
\midrule
Total (core) & -- & All & 197 & 13{,}803+ \\
\bottomrule
\end{tabular}
\caption{Overview of FeedBench, containing 60 short-form videos from 27 content categories.
All core items are grounded in either a temporal interval or a timestamp, instantiated as closed-vocabulary multiple-choice or IoU-scored temporal localization tasks.
$^\dagger$Subject items are currently IoU-only (temporal localization); narrative items are under construction ([WIP]).
Single, Multi, and Ordered denote single-label, multi-label, and ordered-transition items, respectively.}
\label{tab:feedbench_overview}
\end{table}

% ==========================================
\section{Training Process}
\label{sec:training}
\subsection{Supervised Fine-Tuning}
\label{sec:sft}

We conduct SFT on the Qwen3-VL-8B-Instruct~\citep{qwen3vl} backbone to inject domain-specific knowledge for short-video understanding while preserving the model's general-purpose capabilities and instruction-following behavior.
Our principal objective is to endow the model with specialized competencies in temporal video analysis—including camera language understanding, aesthetic assessment, editing evaluation, and temporal grounding—without compromising its proficiency in general multimodal and linguistic tasks.

\subsubsection{Training Objective and Data Composition}

\textbf{SFT as domain-adaptive continued pre-training.}
Rather than treating SFT purely as instruction tuning, we frame it as a form of domain-adaptive continued pre-training that systematically injects specialized knowledge into the foundation model.
By retaining the original Qwen3 response template and system prompt structure throughout training, the model acquires domain-specific capabilities while maintaining its existing instruction-following format and conversational coherence.
This approach ensures that the enhanced short-video understanding abilities integrate seamlessly with the model's pre-existing multimodal reasoning framework.

\textbf{Capability preservation through strategic data mixing.}
To mitigate catastrophic forgetting of general-purpose skills, we adopt a carefully calibrated data mixing strategy.
The training corpus is organized with a deliberate emphasis on domain-specific samples relative to general-purpose ones.
The domain-specific component encompasses diverse short-video analysis tasks, including structured temporal segmentation, multi-dimensional aesthetic scoring, editing quality assessment, and dense video captioning.
The general-purpose component spans visual question answering, OCR, chart understanding, document comprehension, and text-only reasoning tasks.
This heterogeneous composition ensures comprehensive coverage while maintaining the model's versatility across diverse application scenarios.

\textbf{Task-specific system prompt engineering.}
For each constituent dataset, we employ reverse engineering to construct task-specific system prompts that align with the expected response characteristics.
This design serves three complementary purposes: (i) it reinforces instruction-following behavior by providing explicit task context and formatting constraints, (ii) it reduces training loss and perplexity by narrowing the output distribution toward task-appropriate responses, and (iii) it ensures that the model produces sufficiently detailed and well-structured responses even when no explicit system prompt is provided at inference time.
This prompt engineering strategy proves particularly effective for structured output tasks requiring JSON-formatted responses with temporal boundaries, multi-dimensional scores, and hierarchical annotations.

\subsubsection{Training Configuration}

Training is conducted on a distributed GPU cluster using communication- and memory-efficient parallel training together with standard activation and memory optimization techniques.

\textbf{Differential learning rates for multimodal components.}
Following established practices in vision-language model fine-tuning~\citep{liu2024visual,bai2023qwenvl}, we employ differential learning rates across model components: the vision encoder (ViT) and vision-language aligner use a lower learning rate than the language model backbone, preventing excessive drift in pre-trained visual representations while allowing the LLM to adapt more aggressively to the new task distribution.

\textbf{Checkpoint selection via stochastic weight averaging.}
Rather than selecting a single best checkpoint, we apply stochastic weight averaging (SWA) to fuse the final several checkpoints from the training trajectory.
SWA produces a smoother loss landscape and a more robust parameter configuration, which not only improves generalization on held-out benchmarks but also raises the capability ceiling for the subsequent RL stage by providing a stronger and more stable initialization.

\subsection{RLVR (GRPO)}
\label{sec:rlvr}

We apply Group Relative Policy Optimization (GRPO)~\citep{grpo} to refine the SFT checkpoint on perception-oriented tasks.
Unlike reasoning-heavy applications where RL primarily enables test-time scaling through extended chain-of-thought, our setting targets a different benefit: GRPO provides a group-relative reward signal that concentrates the output distribution on correct answers, reducing format instability and label variance even for tasks that do not require multi-step reasoning.

\subsubsection{Motivation}

Two observations motivate the RL stage:
\begin{enumerate}
  \item \textbf{Insensitivity of SFT loss to temporal precision.}
        The token-level cross-entropy objective treats all incorrect predictions equally: a temporal boundary off by 0.1\,s incurs the same loss as one off by 10\,s.
        GRPO's reward function is sensitive to the magnitude of temporal error, directly improving precision on cut-point detection and segment localization.
  \item \textbf{Output distribution drift after domain-adaptive SFT.}
        Large-scale SFT updates both the ViT encoder and the LLM backbone, which can disturb the well-calibrated output distribution of the base model.
        GRPO re-concentrates the distribution toward high-reward responses, effectively re-calibrating the model after the perturbation introduced by SFT.
\end{enumerate}

\subsubsection{Hyperparameters}

RL training is performed on distributed GPU infrastructure with asynchronous rollout and conservative policy updates.
Because our tasks are predominantly perception-oriented rather than reasoning-centric, we use a lightweight configuration relative to chain-of-thought RL systems: a small group size, single gradient step per batch, and cosine learning rate decay.
Key hyperparameters are summarized in Table~\ref{tab:grpo_hparams}.

\begin{table}[htbp]
\centering
\caption{GRPO training hyperparameters.}
\label{tab:grpo_hparams}
\begin{tabular}{@{}ll@{}}
\toprule
\textbf{Parameter} & \textbf{Value} \\
\midrule
Group size $G$                  & 8 \\
Prompts per step                & 390 (13 task types $\times$ 30) \\
Sequences per step (GBS)        & 3,120 ($390 \times 8$) \\
PPO mini-batch size             & 390 (single gradient step) \\
Learning rate                   & $1 \times 10^{-6}$ (cosine $\to 1 \times 10^{-7}$) \\
KL penalty                      & disabled \\
Advantage normalization         & by std (GRPO) \\
Loss aggregation                & seq-mean-token-mean \\
\bottomrule
\end{tabular}
\end{table}

The KL penalty is disabled; over 150 training steps the measured KL between policy and reference remains below 0.3, confirming gradual distribution shift without catastrophic drift.

\subsubsection{Reward Functions}

The reward functions below directly instantiate the SV6D optimization objective (\cref{eq:sv6d_loss}).
The \textbf{IoU} component in temporal grounding rewards corresponds to $\mathcal{L}_{\mathrm{align}}$ (\cref{eq:align_loss}), measuring temporal overlap between predicted and ground-truth shot boundaries.
The \textbf{Label} component corresponds to $\mathcal{L}_{\mathrm{struct}}$ (\cref{eq:struct_loss}), evaluated via the dimension-specific distance functions $d_k$.
The \textbf{Format} component corresponds to $\mathcal{L}_{\mathrm{reg}}$ (\cref{eq:reg_loss}), penalizing out-of-vocabulary labels, missing dimensions, and malformed structure.
For each GRPO rollout, the reward is computed as $r = 1 - \mathcal{L}_{\mathrm{SV6D}}$.
Table~\ref{tab:reward_functions} summarizes the reward formulation per task type.

\begin{table}[htbp]
\centering
\caption{Reward function by task type.}
\label{tab:reward_functions}
\begin{tabular}{@{}ll@{}}
\toprule
\textbf{Task Type} & \textbf{Reward} \\
\midrule
Temporal grounding
  & $0.2\,\mathrm{Format} + 0.4\,\mathrm{IoU} + 0.4\,\mathrm{Label}$ \\
\midrule
Temporal action localization
  & $0.2\,\mathrm{Format} + 0.8\,\mathrm{IoU}$ \\
\midrule
OCR (regular \& handwritten)
  & $\mathrm{EditDist}(\hat{y},\, y^{*})$ \\
\midrule
Chain-of-thought
  & $0.3\,\mathrm{Format} + 0.7\,\mathrm{Judge}$ \\
\bottomrule
\end{tabular}
\end{table}

\subsubsection{RL Dataset and Balancing}
\label{sec:rl_data}

The RL dataset is balanced across domain-video tasks (camera language, cut-point detection, editing, subject, aesthetics), OCR tasks (regular and handwritten), temporal action localization, and a small set of general visual reasoning prompts.
To ensure every gradient step covers all task families, we arrange prompts using a deterministic weighted interleaving scheme rather than random shuffling, guaranteeing that the task-type ratio within any contiguous window closely matches the target distribution.

\subsubsection{Training Dynamics}

\begin{figure*}[t]
\centering
\includegraphics[width=\linewidth]{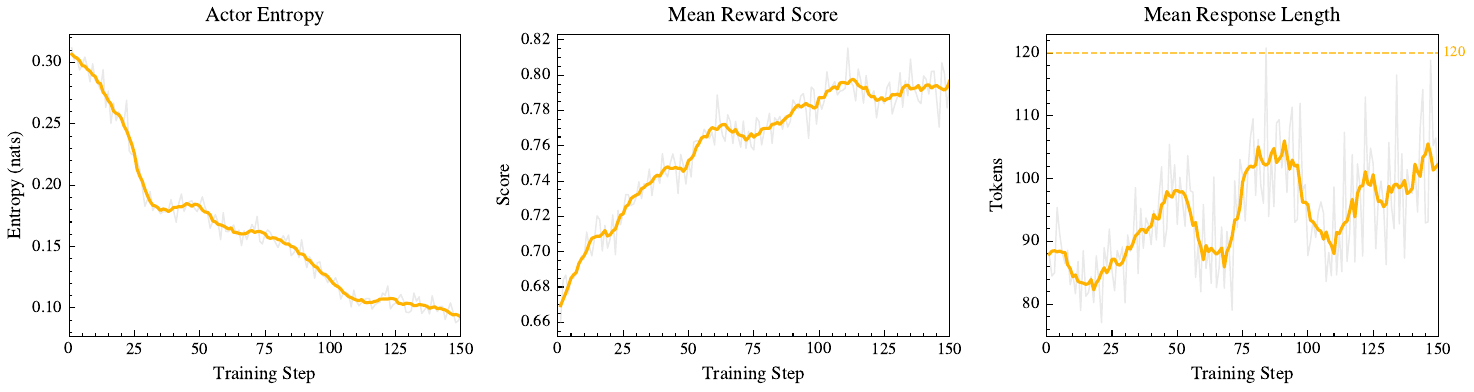}
\caption{GRPO training dynamics over 150 steps.
\textbf{Left}: actor entropy decreases steadily, indicating the policy becomes more confident.
\textbf{Center}: mean reward score rises from $\sim$0.66 to $\sim$0.80.
\textbf{Right}: mean response length remains stable, showing no reward hacking through verbosity.}
\label{fig:grpo_training_curves}
\end{figure*}

Figure~\ref{fig:grpo_training_curves} shows three key metrics over 150 training steps.

Actor entropy decreases steadily over training, indicating the policy becomes more confident.
Mean reward rises consistently, with the steepest gains in the early phase and a plateau in the later steps.
Mean response length remains stable throughout, confirming that the model improves by producing \emph{more accurate} outputs rather than \emph{longer} ones.

\section{Experiments}
\label{sec:eval}

\subsection{FeedBench Results}
% ── Table: FeedBench 6-dimension results ───────────────────────────────────────
\begin{table*}[htbp]
\centering
\caption{\textbf{Performance of \leumvleight on FeedBench.}}
\label{tab:main_results}
\resizebox{\linewidth}{!}{
\begin{tabular}{@{}lcccccc@{}}
\toprule
\textbf{Model} & \textbf{Subject} & \textbf{Aesthetics} & \textbf{Camera Language} & \textbf{Editing} & \textbf{Narrative} & \textbf{Dissemination} \\
\midrule
Qwen3-VL-8B-Instruct$^\dagger$ & 62.3 & 44.0 & 42.0 & 49.2 & -- & 44.9 \\
Qwen3.5-9B~\citep{qwen35}$^\dagger$ & 69.6 & 52.3 & 42.0 & 56.4 & -- & 56.4 \\
\leumvleight                   & \textbf{78.3} & \textbf{64.8} & \textbf{53.0} & \textbf{74.8} & -- & \textbf{71.1} \\
\bottomrule
\end{tabular}
}
\\[0.5em]
\raggedright
\small
$\dagger$ = local reproduction (no official report value). \\
Editing = average of editing logic and cut-point sub-benchmarks. Dissemination = average of dissemination and comments sub-benchmarks. \\
Narrative test set is under construction (WIP); results shown as --. \\
Eval: FPS=4, max 768 frames, max 50K tokens/video. Judge: DeepSeek-V3~\citep{deepseekv3}.
\end{table*}

Table~\ref{tab:main_results} reports per-dimension results on FeedBench.
\leumvleight substantially outperforms both baselines across all evaluated dimensions, with the largest margins on editing (+25.6 pp over Qwen3-VL-8B) and dissemination (+26.2 pp), where correct prediction depends on timeline alignment and expert-sensitive label boundaries rather than coarse scene semantics alone.
On subject framing, \leumvleight scores 78.3, compared to 62.3 for Qwen3-VL-8B-Instruct and 69.6 for Qwen3.5-9B, a gain of +16.0 pp and +8.7 pp respectively.
Notably, these structural gains do not come at the expense of general multimodal capability: \leumvleight remains competitive on Video-MME (70.8), MVBench (70.0), MotionBench (61.6), and MMBench-EN (84.8), supporting our claim that timeline-grounded structural parsing can be added to a compact 8B model without sacrificing broad VLM utility.

\subsection{Open Benchmarks}
% Open benchmark results comparing Leum-VL-8B with Qwen3-VL-8B, Keye-VL-8B, GLM-4.1V-9B, MiniCPM-V-4.5-8B

\begin{table*}[htbp]
\centering
\caption{\textbf{Performance of \leumvleight and other 8B-scale models on open benchmarks.}}
\label{tab:general_results}
\resizebox{\linewidth}{!}{
\begin{tabular}{@{}llccccc@{}}
\toprule
\textbf{Category} & \textbf{Benchmark} & \textbf{\leumvleight} & \textbf{Qwen3-VL-8B}$^1$ & \textbf{Keye-VL-8B Thinking}$^2$ & \textbf{GLM-4.1V-9B Thinking}$^3$ & \textbf{MiniCPM-V-4.5-8B}$^4$ \\
\midrule
\multirow{6}{*}{General VQA}
  & MMBench-EN$_\text{test}$      & \textbf{84.8} & 84.5          & \multirow{2}{*}{92.0}   & 85.8                     & \multirow{2}{*}{84.2} \\
  & MMBench-CN$_\text{test}$      & 83.9          & \textbf{84.7} &                         & 84.7                     & \\
  & HallusionBench               & 56.5          & \textbf{61.1} & 62.7                    & \textbf{63.2}            & 61.2 \\
  & RealWorldQA                  & \textbf{73.2} & 71.5          & 73.5                    & —                        & 72.1 \\
  & MMStar                       & 67.5          & 70.9          & \textbf{80.5}           & 72.9                     & 72.1 \\
  & BLINK                        & \textbf{65.2} & \textbf{69.1} & 54.9$^\ddagger$         & 65.1                     & 42.0$^\ddagger$ \\
\midrule
\multirow{4}{*}{Document \& OCR}
  & OCRBench                     & 85.4          & \textbf{89.6} & 86.6                    & 84.2                     & 89.0 \\
  & DocVQA$_\text{test}$         & \textbf{95.7} & \textbf{96.1} & 93.4$^\ddagger$         & 93.3$^\ddagger$          & 94.7 \\
  & TextVQA$_\text{val}$         & \textbf{85.0} & 82.8$^\ddagger$ & 81.5$^\ddagger$       & 79.6$^\ddagger$          & 82.2 \\
  & ChartQA$_\text{test}$        & 85.3          & \textbf{89.6} & 94.1$^\ddagger$         & 70.0$^\ddagger$          & 87.4 \\
\midrule
\multirow{8}{*}{Video Understanding}
  & Video-MME$_\text{w/o sub.}$  & 70.8          & \textbf{71.4} & 73.0                    & 68.2                     & 67.9 \\
  & MVBench                      & \textbf{70.0} & 68.7          & 56.9$^\ddagger$         & 68.4                     & 60.5$^\ddagger$ \\
  & TempCompass                  & \textbf{74.3} & 74.3$^\ddagger$ & \textbf{75.5}         & 72.3$^\ddagger$          & 72.7$^\ddagger$ \\
  & MotionBench                  & \textbf{61.6} & 56.9$^\ddagger$ & 55.1$^\ddagger$       & 59.0                     & 59.7 \\
  & FAVOR-Bench                  & \textbf{58.9} & 54.1          & —                       & —                        & 56.0 \\
  & LongVideoBench               & 64.6          & 62.4$^\ddagger$ & \textbf{66.0}         & 65.7$^\ddagger$          & 63.9 \\
  & Tomato                       & \textbf{36.7} & 35.7$^\ddagger$ & 33.0$^\ddagger$       & 30.0$^\ddagger$          & 29.8$^\ddagger$ \\
  & Charades-STA$_\text{mIoU}$   & \textbf{59.4} & 56.0          & —                       & —                        & — \\
\bottomrule
\end{tabular}
}
\\[0.5em]
\raggedright
\small
$^\ddagger$ = reported in \citet{molmo2}. \\
$^1$~\citet{qwen3vl}. \quad
$^2$~\citet{keyevl}. \quad
$^3$~\citet{glm41v}. \quad
$^4$~\citet{minicpmv}. \\
Eval: FPS=4, max 768 frames, max 50K tokens/video.
\end{table*}

We evaluate \leumvleight on a comprehensive suite of open benchmarks spanning general visual question answering, document understanding, video comprehension, shot-level analysis, and spatial grounding.
Table~\ref{tab:general_results} summarizes the results.

\textbf{Video understanding.}
\leumvleight achieves strong performance across video benchmarks.
On Video-MME (w/o subtitles), we score 70.8, close to the baseline's official report value of 71.4.
On MotionBench we gain +6.5 pp (61.6 vs.\ 55.1), with the largest improvements on tasks requiring precise temporal localization such as repetition counting and action ordering.
We also improve on MVBench (+1.3 pp), FAVOR-Bench (+4.8 pp), and Charades-STA mIoU (+3.4 pp).

\textbf{Shot-level understanding.}
On RefineShot, we improve overall accuracy by +4.2 pp (56.9 vs.\ 52.7), with the largest gains on Shot Framing (+13.4 pp), Camera Angle (+10.8 pp), and Lighting Type (+6.4 pp)---dimensions that correspond directly to the SV6D schema.

\textbf{Document understanding and OCR.}
\leumvleight remains competitive on text-rich benchmarks: DocVQA$_\text{test}$ (95.7 vs.\ 96.1), TextVQA$_\text{val}$ (+2.8 pp), and OCRBench (85.4, trailing the baseline by 4.2 points).
The gap on OCRBench reflects the training mixture's emphasis on video structure over document-centric tasks.

\textbf{General VQA.}
On MMBench-EN (test) we achieve 84.8 (+0.3 pp over baseline), and improve on RealWorldQA (+1.7 pp), demonstrating that domain-specific training does not degrade general visual reasoning.
Performance drops moderately on MMStar ($-$3.4 pp) and SimpleVQA ($-$4.4 pp), reflecting the expected trade-off toward structured, timeline-grounded outputs.

\textbf{Grounding and multi-image reasoning.}
On spatial grounding (RefCOCO) and multi-image reasoning (BLINK), performance drops moderately, reflecting the training mixture's emphasis on temporal video structure over spatial localization.
We expect this gap to narrow with expanded grounding supervision in future iterations.

\textbf{Summary.}
Overall, the open-benchmark results support the main claim of this report: timeline-grounded structural specialization can be added to a compact 8B model without sacrificing competitive general-purpose multimodal performance.

% ==========================================
\section{Conclusion and Future Work}

We presented \leumvleight, a video-language model specialized in timeline-grounded structural understanding of short-form internet video and commercial creatives.
By formulating video understanding as structured prediction over the SV6D schema---rather than free-form commentary---and combining expert-driven annotation, scalable automated synthesis, and verifiable reinforcement learning, \leumvleight achieves strong performance on both domain-specific and general multimodal benchmarks.

Several directions remain open for future work.
\begin{enumerate}
  \item \textbf{Continued pre-training on domain corpora.} Injecting larger-scale video production and operations knowledge through continued pre-training would further ground the model in the professional vocabulary and reasoning patterns of cinematography, directing, and platform-native content creation.
  \item \textbf{End-to-end agentic capabilities.} Strengthening training on RL and agent-loop objectives would enable the model to support end-to-end workflows spanning creative planning, shoot preparation, reference-based creation, post-production editing, and distribution operations.
  \item \textbf{Audio-visual language understanding.} Extending the architecture to incorporate an audio encoder would allow the model to reason jointly over BGM, rhythmic beat alignment, vocal tone, and the interplay between audio and visual expression---an important layer of meaning in short-form video that the current model does not address.
\end{enumerate}

\clearpage
\bibliography{references,datasets_bibtex}
\bibliographystyle{colm2024_conference}

\appendix
\renewcommand{\thesection}{\Alph{section}}
\titleformat{\section}{\large\bfseries}{Appendix \thesection}{1em}{}
\clearpage
\section{Benchmarks}
\label{app:benchmarks}

We evaluate \leumvleight on a range of public benchmarks covering general visual question answering, document understanding, video understanding, visual grounding, and our in-house short-video benchmark FeedBench. Below we provide a description of each benchmark used.

\begin{itemize}[itemindent=0pt, labelsep=4pt, leftmargin=*]

\item \textbf{General Visual Question Answering:}
\begin{itemize}
  \item \textbf{MMBench}~\citep{mmbench}: A large-scale multi-ability VQA benchmark with English (EN) and Chinese (CN) splits. We report test-set accuracy (submitted to the official leaderboard). Evaluation uses circular evaluation (each question asked four times in shuffled order; all must be correct to count), which strongly tests answer stability.
  \item \textbf{MMStar}~\citep{mmstar}: A carefully curated benchmark of 1,500 challenging visual questions designed to minimize language-only solvability.
  \item \textbf{RealWorldQA}~\citep{realworldqa}: A benchmark of real-world spatial and scene understanding questions sourced from vehicle cameras and everyday environments.
  \item \textbf{HallusionBench}~\citep{hallusionbench}: A benchmark specifically designed to probe visual hallucination and language-prior bias in VLMs.
  \item \textbf{BLINK}~\citep{blink}: A multi-image perception benchmark requiring cross-image comparison, spatial reasoning, and visual correspondence.
  \item \textbf{SimpleVQA}~\citep{simplevqa}: A factual VQA benchmark testing world knowledge grounded in visual evidence.
\end{itemize}

\item \textbf{Document Understanding:}
\begin{itemize}
  \item \textbf{OCRBench}~\citep{ocrbench}: A comprehensive OCR evaluation covering text recognition, scene-text VQA, document-oriented VQA, key information extraction, and handwritten mathematical expression recognition. We report the normalized final score (out of 100).
  \item \textbf{DocVQA}~\citep{docvqa}: Document visual question answering evaluated with ANLS (Average Normalized Levenshtein Similarity). We report the test split.
  \item \textbf{TextVQA}~\citep{textvqa}: Scene-text VQA requiring reading and reasoning over text embedded in natural images. We report val-set accuracy.
  \item \textbf{ChartQA}~\citep{chartqa}: Chart understanding benchmark evaluated with relaxed accuracy. We report test-set results.
\end{itemize}

\item \textbf{Video Understanding:}
\begin{itemize}
  \item \textbf{Video-MME}~\citep{videomme}: A comprehensive video QA benchmark spanning short, medium, and long videos. We evaluate without subtitles (w/o sub.) and report overall accuracy.
  \item \textbf{MVBench}~\citep{mvbench}: A comprehensive video QA benchmark covering 20 temporal understanding tasks, including action recognition, scene transition, and object interaction.
  \item \textbf{TempCompass}~\citep{tempcompass}: A temporal reasoning benchmark testing fine-grained understanding of event order, duration, and speed in videos.
  \item \textbf{MotionBench}~\citep{motionbench}: A motion-centric video understanding benchmark with six sub-tasks: Motion Recognition, Location-related Motion, Camera Motion, Motion-related Objects, Action Order, and Repetition Count. We report the dev-set average.
  \item \textbf{FAVOR-Bench}~\citep{favorbench}: A fine-grained video motion understanding benchmark evaluating temporal perception across diverse motion categories.
  \item \textbf{LongVideoBench}~\citep{longvideobench}: A benchmark for long-context interleaved video-language understanding, testing comprehension over extended video sequences.
  \item \textbf{Tomato}~\citep{tomato}: A visual temporal reasoning benchmark assessing multimodal models' ability to understand temporal order, duration, and causal relationships in video.
\end{itemize}

\item \textbf{Shot and Camera Understanding:}
\begin{itemize}
  \item \textbf{RefineShot}~\citep{refineshot}: A cinematography benchmark covering eight shot-level attributes: lens size (LS), lighting type (LT), lighting condition (LC), shot framing (SF), shot size (SS), camera angle (CA), shot composition (SC), and camera movement (CM). Evaluation uses step-by-step prompting with consistency checking to ensure reasoning and answer alignment.
  \item \textbf{Charades-STA}~\citep{charadessta}: A temporal grounding benchmark requiring localization of natural-language described events in video. We report mean IoU (mIoU) on the test set.
\end{itemize}

\item \textbf{Visual Grounding:}
\begin{itemize}
  \item \textbf{RefCOCO/+/g}~\citep{refcoco}: Referring expression comprehension benchmarks. We report the average accuracy (Acc@0.5 IoU) across RefCOCO val, testA, testB, RefCOCO+ val, testA, testB, and RefCOCOg val, test.
\end{itemize}

\end{itemize}

% Demo conversation box styles
\newcommand{\imgcaption}[1]{%
  \par\smallskip
  {\centering\small\color{gray!70!black}#1\par}%
  \smallskip
}

\tcbset{
  demobox/.style={
    fontupper=\sffamily,
    fonttitle=\bfseries\small\sffamily,
    toptitle=4pt, bottomtitle=4pt,
    left=6pt, right=6pt, top=4pt, bottom=4pt,
    boxsep=0pt
  },
  userbox/.style={
    demobox,
    colback=blue!5, colframe=blue!40,
    title={User},
    before skip=6pt, after skip=2pt
  },
  assistantbox/.style={
    demobox,
    colback=gray!5, colframe=gray!50,
    title={Assistant},
    before skip=2pt, after skip=6pt
  }
}

\section{Demonstrations}
\label{app:demonstrations}

We present three qualitative demonstrations of \leumvleight on real-world internet videos, showcasing its ability to perform complex video understanding tasks including editing strategy analysis, structured information extraction, and cinematographic evaluation.

% ─────────────────────────────────────────────
\subsection{Viral Editing Strategy Analysis}
\label{app:demo:editing}

\begin{tcolorbox}[userbox]
If the video were to be trimmed and re-edited to approximately 20 seconds, which segments should be retained to maximize its viral potential while maintaining narrative integrity? Please provide the rationale.

\medskip\noindent\includegraphics[page=1, width=\linewidth, trim=18pt 96pt 18pt 96pt, clip]{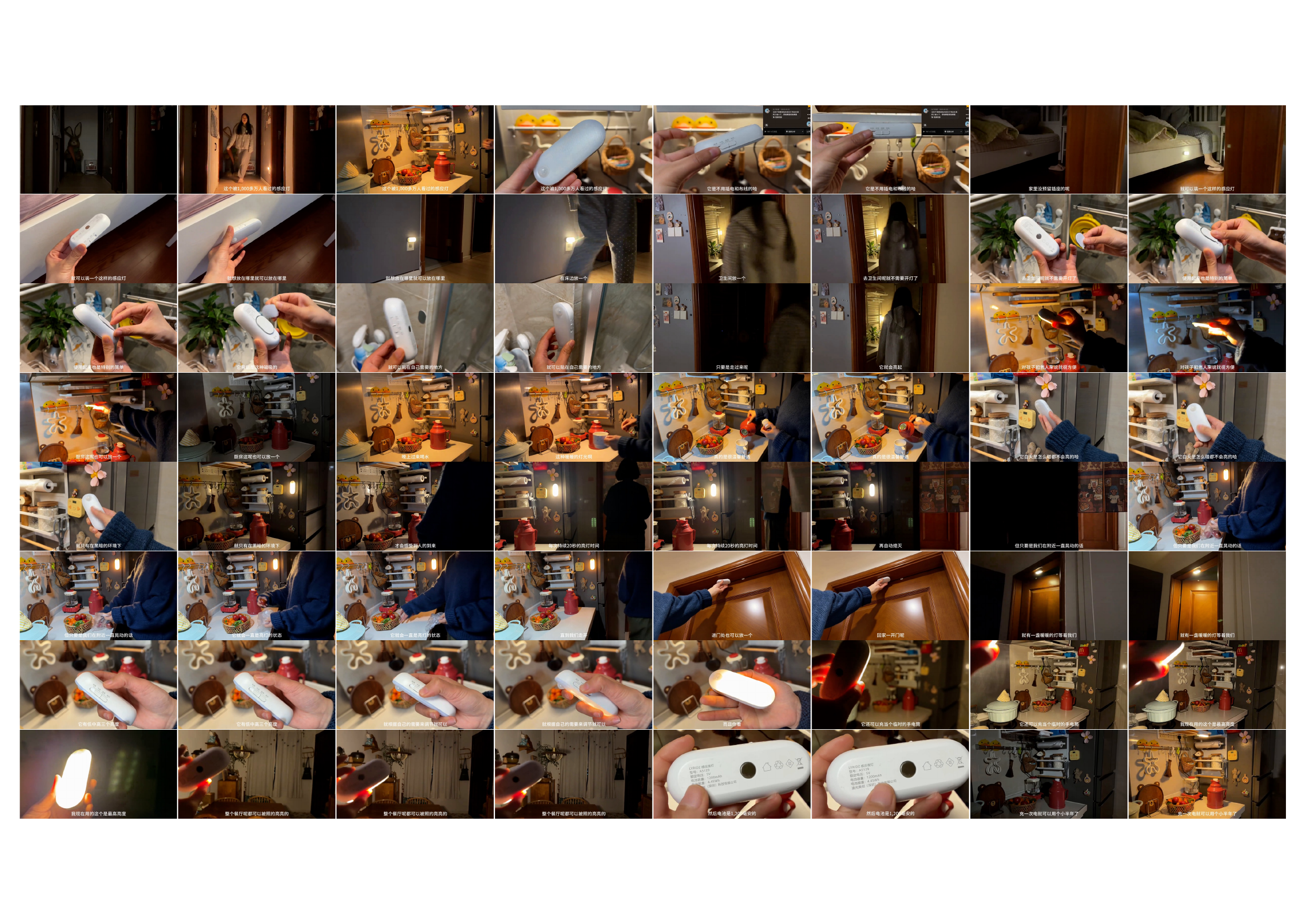}
\imgcaption{Input Video 00:00:58}
\end{tcolorbox}

\begin{tcolorbox}[assistantbox]
Retained: [00:00.0--00:04.4] [00:09.5--00:13.3] [00:17.9--00:21.8] [00:26.8--00:31.7] [00:43.4--00:47.2] [00:54.0--00:57.9]

Discarded: [00:04.4--00:17.9] [00:21.8--00:39.3] [00:39.3--00:49.8] [00:49.8--00:54.0]

Rationale: The video unfolds through a linear structure of ``Pain Point Introduction -- Scenario Demonstration -- Core Features -- Value Reinforcement.'' The opening immediately leverages a high-engagement endorsement of ``over 10 million viewers'' and emphasizes the core selling point of ``no wiring or plugging required'' to rapidly capture attention. This is followed by a fast-paced montage across four typical life scenarios---bedside, bathroom, kitchen, and entryway---to intuitively demonstrate the convenience of ``magnetic installation'' and ``motion-sensor lighting,'' highlighting practicality for demographics like the elderly and children. Subsequently, the edit focuses on core functionalities such as ``adjustable brightness levels'' and ``flashlight mode'' to showcase product versatility. Finally, it concludes with hardcore specifications of a ``1200mAh battery and nearly six months of endurance'' to build trust through data. This compact ``Pain Point -- Scenario -- Feature -- Parameter'' narrative ensures high information density and aligns with the ``Golden 3 Seconds'' rule of short-form video, effectively enhancing completion and conversion rates.

\medskip\noindent\includegraphics[page=2, width=\linewidth, trim=18pt 96pt 17pt 95pt, clip]{figures/demo_pictures}
\imgcaption{Retained Time Segments}
\end{tcolorbox}

% ─────────────────────────────────────────────
\subsection{Structured Step Extraction with Causal Dependencies}
\label{app:demo:steps}

\begin{tcolorbox}[userbox]
Extract the core executable steps from the video. Requirements: Distinguish between ``visibly demonstrated actions'' and ``verbally stated instructions,'' provide the causal dependencies between the steps, and output the result in JSON format.

\medskip\noindent\includegraphics[page=3, width=\linewidth, trim=18pt 98pt 17pt 97pt, clip]{figures/demo_pictures}
\imgcaption{Input Video 00:00:38}
\end{tcolorbox}

\begin{tcolorbox}[assistantbox]
\begin{lstlisting}[language={}, basicstyle=\ttfamily\footnotesize, frame=none]
{
  "visibly demonstrated actions": [
    {
      "step": "Depicting subjects' awkwardness and unease in public settings",
      "time_range": ["00:00:00.0", "00:00:15.0"],
      "details": "By showcasing the awkwardness and unease of various
        individuals in public, it implies the privacy leakage issue of
        health data being known by others.",
      "evidence": "The man who hasn't brushed his teeth since Tuesday
        enters the room. Said Hillary, who's only taken 347 steps today.
        And then there's Bill, who woke up to pee twice last night."
    },
    {
      "step": "Demonstrating the sensitivity of health data",
      "time_range": ["00:00:15.0", "00:00:21.0"],
      "details": "Using specific examples to illustrate the sensitivity
        of health data, such as knowing someone was gassy yesterday.",
      "evidence": "How do they know so much about us? Your health data
        says a lot about you. That's how I know Kevin was gassy yesterday."
    },
    {
      "step": "Showcasing the privacy protection of health data",
      "time_range": ["00:00:21.0", "00:00:28.5"],
      "details": "By displaying the iPhone Health App interface, it
        emphasizes its privacy protection features to reassure users.",
      "evidence": "Your health data is personal. The Health app helps
        keep it private. Privacy. That's iPhone."
    }
  ],
  "verbally stated instructions": [
    {
      "step": "Emphasizing the personal privacy of health data",
      "time_range": ["00:00:16.5", "00:00:18.5"],
      "details": "Using voiceover to underscore the personal privacy of
        health data, resonating with the audience.",
      "evidence": "Your health data says a lot about you."
    },
    {
      "step": "Stating the privacy protection measures for health data",
      "time_range": ["00:00:23.0", "00:00:28.5"],
      "details": "Explicitly stating the privacy features via voiceover
        while showing the Health App interface to build user trust.",
      "evidence": "Your health data is personal. The Health app helps
        keep it private. Privacy. That's iPhone."
    }
  ]
}
\end{lstlisting}
\end{tcolorbox}

% ─────────────────────────────────────────────
\subsection{Opening Hook Identification and Cinematographic Ranking}
\label{app:demo:hooks}

\begin{tcolorbox}[userbox]
Identify the three segments most suitable as ``opening hooks,'' and analyze as well as rank them based on content structure, visual aesthetics, and shot scales.

\medskip\noindent\includegraphics[page=4, width=\linewidth, trim=18pt 239pt 17pt 96pt, clip]{figures/demo_pictures}
\imgcaption{Input Video 00:00:25}
\end{tcolorbox}

\begin{tcolorbox}[assistantbox]
\textbf{1. Segment 1: Surreal Opening at the Supermarket Checkout (00:00--00:05)}
\begin{itemize}[leftmargin=*, itemsep=2pt, topsep=2pt]
  \item \textit{Content Structure:} This segment utilizes the everyday setting of a supermarket checkout, creating strong visual contrast and a sense of surrealism through hanging apples, bizarre scanning motions, and a final close-up of a giant floating apple. It rapidly captures the audience's attention, serving as an effective opening hook.
  \item \textit{Visual Aesthetics:} Employs a centered composition, with the subject and shopping basket positioned in the middle of the frame, flanked by symmetrically distributed shelves. The lighting is uniform indoor artificial light with medium color saturation and an overall warm tone, creating a cozy yet slightly dreamlike atmosphere.
  \item \textit{Shot Scale:} A medium shot is used, framing the subject from the waist up to the top of the head. This fully displays upper-body movements and facial expressions while retaining some background information from the shelves.
  \item \textit{Rationale:} Ranked first because the sharp contrast between ``daily life'' and ``surreal elements'' breaks the viewer's visual expectations in the shortest time possible, stimulating immediate curiosity.
\end{itemize}

\textbf{2. Segment 2: Close-up and Slicing of the Floating Apple (00:05--00:06)}
\begin{itemize}[leftmargin=*, itemsep=2pt, topsep=2pt]
  \item \textit{Content Structure:} Following the initial visual impact of the opening, the dynamic footage of the apple being sliced by a supermarket barcode scanner demonstrates a playful and creative visual effect, further intensifying the visual appeal.
  \item \textit{Visual Aesthetics:} Employs a centered composition with a giant red apple occupying the center of the frame against a blurred background. Light focuses on the apple's surface to create specular highlights; the colors are vivid with high contrast, making it highly visually appealing.
  \item \textit{Shot Scale:} A close-up shot is used, where the giant red apple fills nearly the entire frame, leaving only minimal background context to emphasize the object's texture and detail.
  \item \textit{Rationale:} Ranked second as it sustains the suspense of the opening, objectifying the abstract concept of ``cuteness'' through specific dynamic details (the slicing), thereby maintaining viewer interest.
\end{itemize}

\textbf{3. Segment 3: Character Hat-removal Transformation (00:07--00:09)}
\begin{itemize}[leftmargin=*, itemsep=2pt, topsep=2pt]
  \item \textit{Content Structure:} Through the action of removing a hat, the character completes an identity shift from a ``working state'' to a ``home state.'' This sets the stage for demonstrating the comfort of loungewear and serves as a key node for plot progression.
  \item \textit{Visual Aesthetics:} Employs a centered composition with the character's face in the middle, set against a background of blurred cartoon plushies. The lighting is soft and uniform with warm colors, creating a comfortable and healing domestic atmosphere.
  \item \textit{Shot Scale:} A medium close-up is used, framing the character from the head to below the chest. It focuses on facial expressions and upper-body movements, with the background blurred to highlight the subject.
  \item \textit{Rationale:} Ranked third because while it achieves a narrative transition, its visual impact is relatively weaker than the first two segments, acting more as an emotional bridge.
\end{itemize}
\end{tcolorbox}

\end{document}